\documentclass[iop]{emulateapj}
\slugcomment{Accepted to AJ}

\shorttitle{Nonmetastable Ammonia Masers in NGC~7538}
\shortauthors{Hoffman \& Kim}

\begin{document}

\title{New Maser Emission from Nonmetastable Ammonia in NGC~7538. II. Green Bank Telescope Observations Including Water Masers}

\author{Ian M.\ Hoffman}
\affil{St.\ Paul's School, Concord, NH 03301}
\email{ihoffman@sps.edu}

\and

\author{Stella Seojin Kim\altaffilmark{1}}
\affil{St.\ Paul's School, Concord, NH 03301}
\altaffiltext{1}{present address: Massachusetts Institute of Technology, 77 Massachusetts Avenue, Cambridge, MA 02139}

\begin{abstract}
We present new maser emission from $^{14}$NH$_3$ (9,6) in NGC~7538.
Our observations include the known spectral features near $v_{\rm LSR} = -60\,{\rm km}\,{\rm s}^{-1}$ and $-57\,{\rm km}\,{\rm s}^{-1}$ and several more features extending to $-46\,{\rm km}\,{\rm s}^{-1}$.
In three epochs of observation spanning two months we do not detect any variability in the ammonia masers, in contrast to the $>$10-fold variability observed in other $^{14}$NH$_3$ (9,6) masers in the Galaxy over comparable timescales.
We also present observations of water masers in all three epochs for which emission is observed over the velocity range $-105\,{\rm km}\,{\rm s}^{-1} < v_{\rm LSR} < -4\,{\rm km}\,{\rm s}^{-1}$, including the highest velocity water emission yet observed from NGC~7538.
Of the remarkable number of maser species in IRS~1, H$_2$O and, now, $^{14}$NH$_3$ are the only masers known to exhibit emission outside of the velocity range $-62\,{\rm km}\,{\rm s}^{-1} < v_{\rm LSR} < -51\,{\rm km}\,{\rm s}^{-1}$. 
However, we find no significant intensity or velocity correlations between the water emission and ammonia emission.
We also present a non-detection in the most sensitive search to date toward any source for emission from the CC$^{32}$S and CC$^{34}$S molecules, indicating an age greater than $\approx 10^4$~yr for IRS~1-3.
We discuss these findings in the context of embedded stellar cores and recent models of the region.
\end{abstract}


\section{Introduction}

\begin{figure*}
%\epsscale{.80}
%\plottwo{f1a.eps}{f1b.eps}
\begin{minipage}[b]{0.48\textwidth}
\noindent
\centering
\includegraphics*[angle=270,scale=0.43]{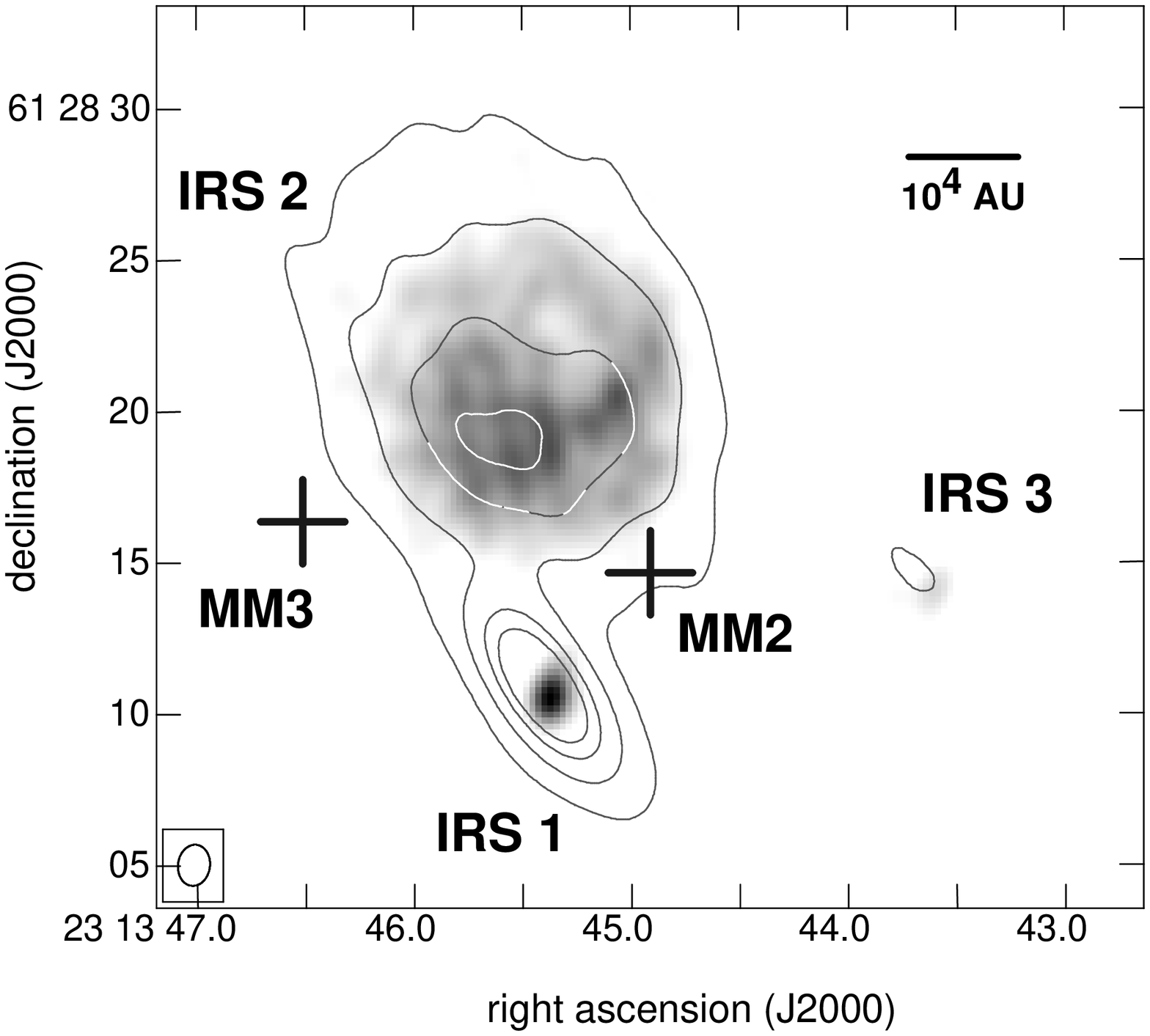}
\end{minipage}
%\hspace{0.50cm}
\hfill
\begin{minipage}[t]{0.48\textwidth}
\noindent
\centering
\includegraphics*[angle=270,scale=0.40]{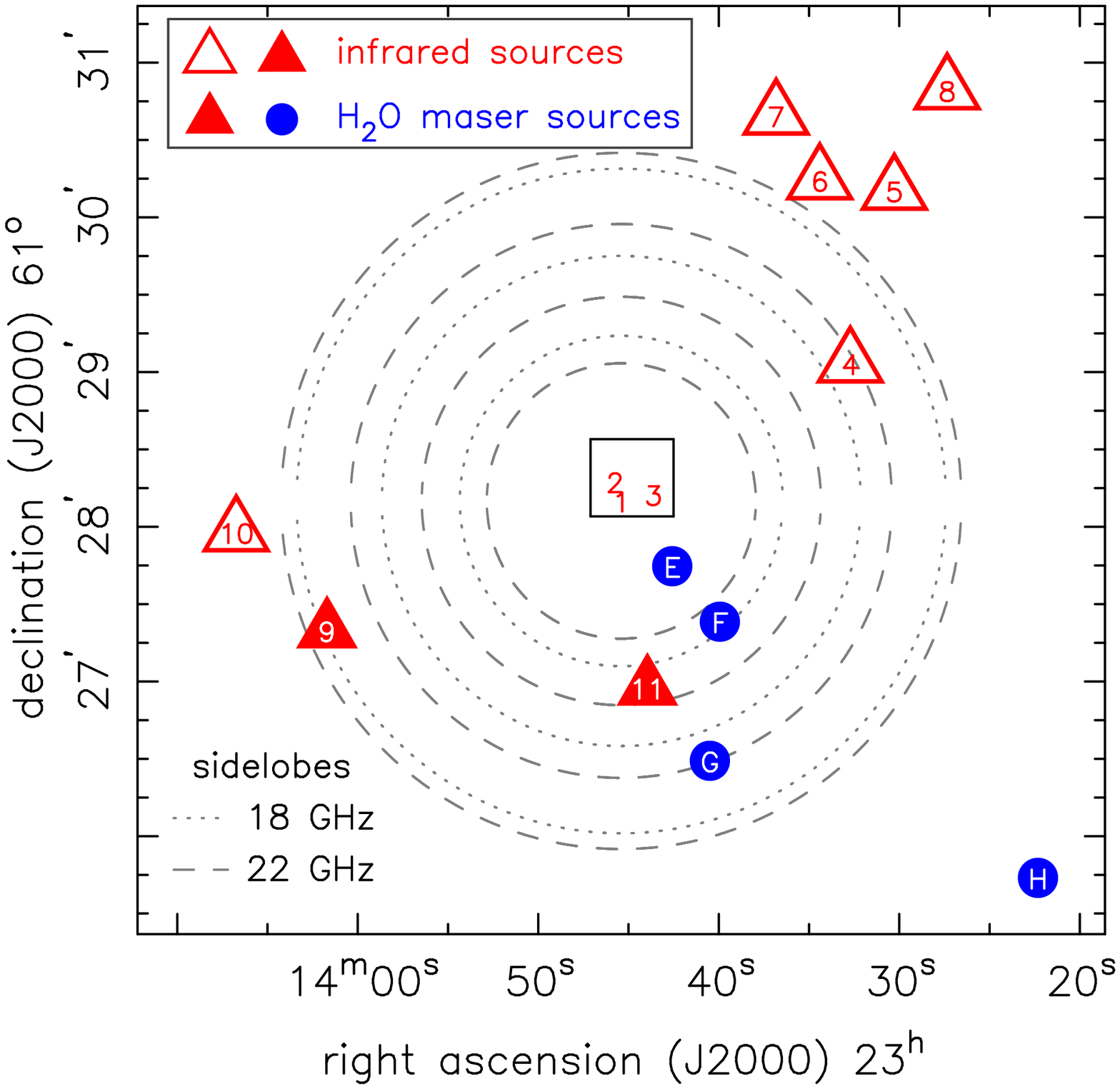}
\end{minipage}
\caption{
{\it (a)} At left are images of the continuum emission from the IRS~1-3 region.
The greyscale is the VLA `B'-array image at 4.8~GHz (6~cm) from Hoffman et al.\ (2003).
The contours are the EVLA `DnC'-array image at 18.5~GHz (1.6~cm) from Hoffman \& Kim (Paper~I).
The crosses mark the locations of the millimeter sources MM2 and MM3 from Qiu et al.\ (2011) that have coincident water maser emission, as discussed in \S\ref{waterResults}.
The position registration among these data sets is uncertain by approximately 500~milliarcseconds.
The greyscale flux ranges from 5.5 to 56.5~${\rm mJy}\,{\rm beam}^{-1}$.
The contour levels are 15, 52, 104, and 150~${\rm mJy}\,{\rm beam}^{-1}$.
The beam for the 6-cm image is $1.4 \times 1.1$~arcseconds (shown at lower left).
The beam for the 1.6-cm image is $4.6 \times 1.9$~arcseconds.
The length of the line at the upper right represents an image scale of $10^4$~AU for a distance to NGC~7538 of 2.65~kpc.
%The width of the area represented in panel {\it (b)} is nearly 4~pc.
{\it (b)} At right is a plot of the positions of the infrared sources (triangles, Werner et al.\ 1979) and water maser positions (circles) with filled symbols indicating associated water maser emission observed by Kameya et al.\ (1990).
The triangle symbols have been omitted from IRS 1, 2, and 3 for clarity.
The rectangle enclosing IRS~1-3 at the center is the boundary of the image in {\it (a)}.
Overlaid on the symbols is a plot of the beam response of the Green Bank Telescope for our observations.
The dotted and dashed lines are sidelobe peaks in the Airy pattern.
The first three 18-GHz sidelobes are shown with dotted lines and the first four 22-GHz sidelobes are shown with dashed lines.
For these observations, the main beam was centered on IRS~1 and has a full width at half maximum of approximately 30~arcseconds, completely enclosing the imaged area shown in {\it (a)}.
The sidelobe levels are given in \S2.
IRS 7 and 8 are suggested to be unassociated with the NGC~7538 complex (e.g., Zheng et al.\ 2001).
\label{6cm}}
\end{figure*}

The sources in NGC~7538 comprise a well studied complex of forming stars at a distance of 2.7~kpc (Moscadelli et al.\ 2009).
Eleven infrared sources have been identified (Werner et al.\ 1979), three of which (IRS~1, 9, and 11) have associated water and hydroxyl maser emission (Kameya et al.\ 1990; Hutawarakorn \& Cohen 2003) and one of which (IRS~1) hosts the largest single collection of observed maser species in the Galaxy (e.g., Galv\'an-Madrid et al.\ 2010).
In addition to illuminating IRS objects, the water masers are observed at several additional locations in the complex.
Furthermore, the water masers are observed over a much wider range in velocity ($\approx 100\,{\rm km}\,{\rm s}^{-1}$) than all other maser species, which have only ever been observed over the limited range $-62\,{\rm km}\,{\rm s}^{-1} < v_{\rm LSR} < -51\,{\rm km}\,{\rm s}^{-1}$ near the apparent systemic velocity of $v_{\rm LSR} \approx -58\,{\rm km}\,{\rm s}^{-1}$.
Qiu et al.\ (2011), based on observations at submillimeter wavelengths, have recently reported several additional sources of continuum emission near IRS~1 in NGC~7538 that are described as embedded stellar cores (see also Akabane et al.\ 1992; Kawabe et al.\ 1992).
Two of these sources, MM2 and MM3, have gas masses comparable to that of the central star in IRS~1 ($\approx 20\,M_\odot$) and have positions in common with water masers.
The ongoing massive star formation near IRS~1 is described by Qiu et al.\ (2011) as the ``Trapezium'' of NGC~7538.
In Figure~\ref{6cm} we show both the IRS~1-3 region and the larger complex including all other IRS objects.

Surcis et al.\ (2011) have formed the most recent model of IRS~1, incorporating new observations of water and methanol masers with other line (e.g., Qiu et al.\ 2011) and continuum (e.g., Sandell et al.\ 2009) observations in order to delineate an outflow, torus, and circumstellar disk.
This model can be used to explain the line emission in the narrow velocity range $-62\,{\rm km}\,{\rm s}^{-1} < v_{\rm LSR} < -51\,{\rm km}\,{\rm s}^{-1}$ but it does not address the details of the outflow or other sources presumably responsible for the ``high''-velocity line emission outside of this narrow range, typified by the water masers.

The $(J,K) = (9,6)$ $^{14}$NH$_3$ maser was discovered serendipitously by Madden et al.\ (1986) toward four of 17 Galactic star-forming regions surveyed: W51, W49, DR21(OH), and NGC~7538.
In 2010, we observed the maser for the first time since its discovery, using the EVLA (Hoffman \& Kim 2011, hereafter Paper~I).
We found (9,6) maser emission at new velocities, covering the range $-60\,{\rm km}\,{\rm s}^{-1} < v_{\rm LSR} < -56\,{\rm km}\,{\rm s}^{-1}$, associated in position with IRS~1, and consistent with the kinematics of the model of Surcis et al.\ (2011).
Elsewhere in the Galaxy, nonmetastable ($J>K$) ammonia masers are known to be variable (W49: Madden et al.\ 1986; W51: Wilson, Henkel, \& Johnston 1990).
For example, Wilson \& Henkel (1988) observed intensity variability of the maser in W51 by a factor of 20 over a timescale of 10~months.
In a search for common variability of ammonia and water masers and for other new constraints on this well studied star-forming complex, we have undertaken new $K$-band radio observations of NGC~7538.

\newpage
\section{Observations\label{obs}}

\begin{deluxetable}{l c l}
%\tablecolumns{3}
\tablewidth{0pt}
\tablecaption{Observed Transitions\label{rest}}
\tablehead{
\colhead{molecule} & \colhead{transition} & \colhead{$\nu_{\rm rest}$} \\
\colhead{} & \colhead{} & \colhead{(MHz)}}
\startdata
$^{14}$NH$_3$ & $9,6$               & 18499.390 \\
CC$^{34}$S    & $2_1 - 1_0$         & 21930.476 \\
H$_2$O        & $6_{1,6} - 5_{2,3}$ & 22235.120 \\
CC$^{32}$S    & $2_1 - 1_0$         & 22344.030 \\ [-3mm]
\enddata
\tablecomments{The notation for the transitions is $J,K$ for ammonia, $J_N$ for thioxoethenylidene, and $J_{Ka,Kc}$ for water.}
\end{deluxetable}

We observed three separate epochs using the Green Bank Telescope of the NRAO\footnote{The National Radio Astronomy Observatory is a facility of the National Science Foundation operated under cooperative agreement by Associated Universities, Inc.} on 2010 November 24, 2010 December 8, and 2011 January 22.
The November and December epochs each had an integration time on the science target of approximately 30~minutes while the January epoch had a target integration time of approximately 1~hour.
We used the first-generation two-beam $K$-band receiver system for nodding observations between the beams using the ``lower'' (18-22~GHz) horns and polarizers (Srikanth 2009).
The polarizers are approximately 20\% less sensitive at 18.5~GHz and 22.3~GHz than at their peak at 20~GHz (S.\ Srikanth, 2011, private communication).
Dual circular polarizations were recorded.
We employed four spectral windows and three-level sampling in the NRAO autocorrelation spectrometer (ACS) yielding 16384 channels across a 12.5-MHz bandwidth, corresponding to a usable velocity coverage of $-140\,{\rm km}\,{\rm s}^{-1} < v_{\rm LSR} < 30\,{\rm km}\,{\rm s}^{-1}$ and a channel spacing of $0.010\,{\rm km}\,{\rm s}^{-1}$.
The rest frequencies and molecular transitions used to center the four windows on $v_{\rm LSR} = -60\,{\rm km}\,{\rm s}^{-1}$ are summarized in Table~\ref{rest}.
Amplitude calibration, focusing, and pointing were based on observations of 3C48, 3C123, and J2322+509.
Typical {\it rms} noise levels in line-free channels were 23~mJy for the November and December epochs and 13~mJy for the January epoch.

In Figure~\ref{6cm} is shown the observed region.
In Figure~\ref{6cm}b is shown a plot of the positions of the infrared (IRS) sources (Werner et al.\ 1979) and water masers (Kameya et al.\ 1990) in NGC~7538 overlaid with the beam and sidelobes of the GBT at both 18~GHz and 22~GHz.
The relative amplitude of the first four 22-GHz sidelobes compared with the main beam are, respectively, $-29.9$~dB, $-30.9$~dB, $-35.6$~dB, and $-38.0$~dB, and for the first three 18-GHz sidelobes: $-28.6$~dB, $-30.2$~dB, and $-33.0$~dB (Srikanth 2009).
There is negligible antenna sensitivity in the nulls of the beam pattern between the plotted sidelobe peaks.

\section{Results and Discussion}

\subsection{NH$_3$\label{results}}

\begin{figure*}
%\epsscale{.80}
\centering
\includegraphics*[angle=270,scale=0.67]{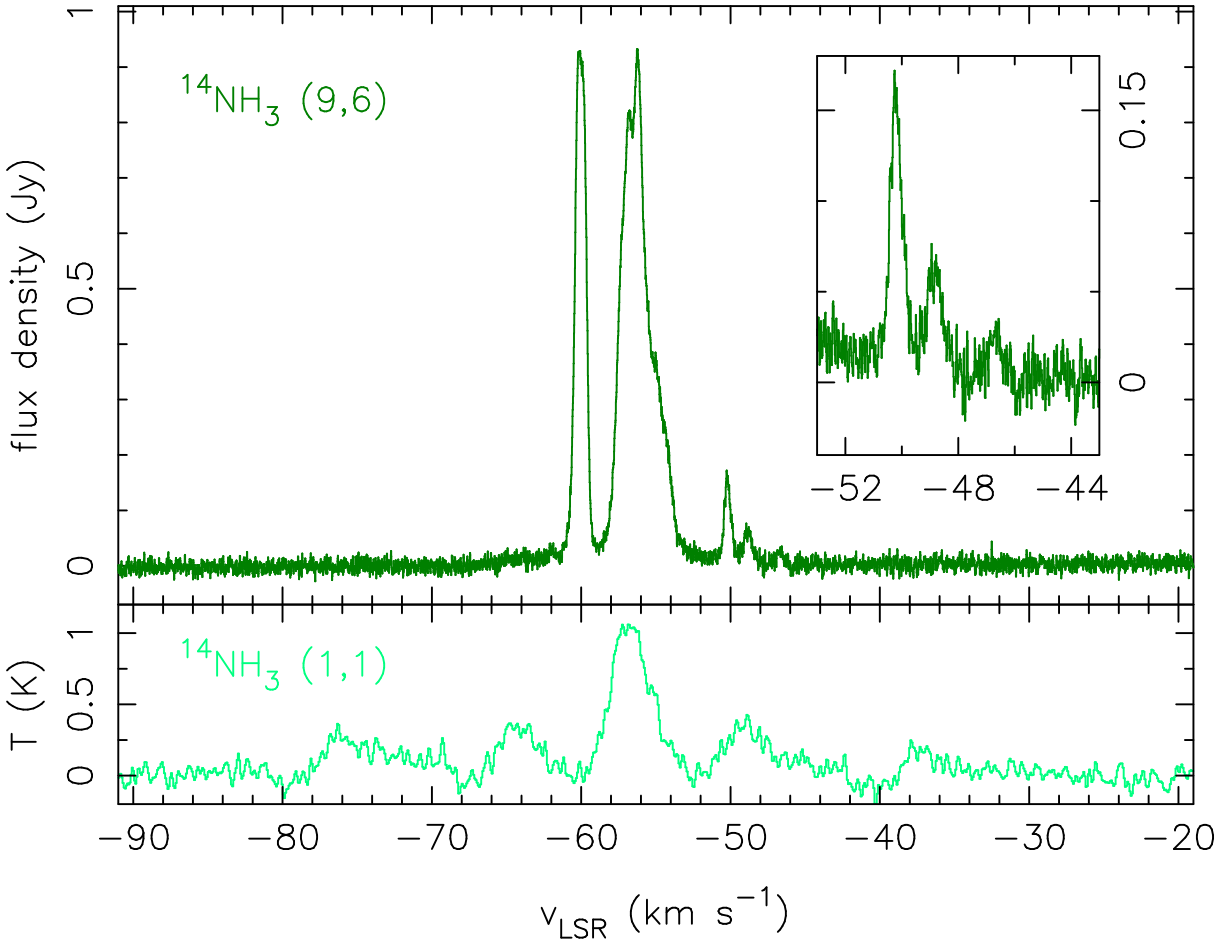}
%\plotone{spectrum.eps}
\caption{
{\it (top)} GBT spectrum of the $^{14}$NH$_3$ $(9,6)$ masers in NGC~7538 from 2011 January 22.
(The 2010 November and December epochs have no significant differences from the epoch shown, as discussed in \S\ref{results}.)
The fitted properties of the emission features are summarized in Table~\ref{table}.
{\it (inset)} A magnification of the three weakest $(9,6)$ emission features.
{\it (bottom)} The thermal emission from the (1,1) transition observed using the GBT by Urquhart et al.\ (2011) as part of the RMS survey.
Their (1,1), (2,2), and (3,3) observations are discussed in \S\ref{results}.
\label{gbt}}
\end{figure*}

\begin{deluxetable}{l l l c}
\tablecolumns{4}
\tablewidth{0pt}
\tablecaption{Fitted Spectral Properties of Ammonia Emission\label{table}}
\tablehead{
\colhead{$v_{\rm LSR}$} & \colhead{$S$\tablenotemark{$\dagger$}} & \colhead{$\Delta{v}_{\rm FWHM}$} & \colhead{Paper~I} \\
\colhead{(${\rm km}\,{\rm s}^{-1}$)} & \colhead{(Jy)} & \colhead{(${\rm km}\,{\rm s}^{-1}$)} & \colhead{label}
}
\startdata
$-$60.16(2)  & 0.94(2)  & 0.64(2) & `SE' \\
$-$59.72(2)  & 0.44(6)  & 0.44(2) & `SE' \\
$-$56.93(1)  & 0.49(1)  & 1.16(2)\tablenotemark{a} & `NW' \\
$-$56.14(1)  & 0.34(1)  & 0.55(1)\tablenotemark{a} & - \\
$-$55.78(1)  & 0.44(1)  & 2.84(1)\tablenotemark{a} & \tablenotemark{b} \\
$-$50.20(2)  & 0.15(1)  & 0.58(1) & \tablenotemark{b} \\
$-$48.85(1)  & 0.057(2) & 0.82(3) & \tablenotemark{b} \\
$-$46.71(3)  & 0.023(2) & 0.65(7) & \tablenotemark{b} \\ [-3mm]
\enddata
\tablecomments{The number in parentheses is the uncertainty in the final digit.}
\tablenotetext{$\dagger$}{The absolute intensity calibration of the observations is precise within $\approx 10$\% while the higher precision quoted here is appropriate for relative comparisons among the features.}
\tablenotetext{a}{The fits to these features are likely unreliable due to blending of several narrower features.}
\tablenotetext{b}{The bandwidth of the EVLA observations presented in Paper~I did not cover these velocities.}
\end{deluxetable}

Ammonia maser emission was detected in each of the three epochs.
In Figure~\ref{gbt} and Table~\ref{table} are shown the results from 2011 January 22.
We do not detect any variability among any of the epochs.
Aside from multiplicative scaling ($\pm 10$\%) representative of typical calibration uncertainties, the spectra from the three epochs have no significant differences.
Although the absolute calibration is uncertain at a $\approx$10\% level, the relative magnitude of the amplitude parameters listed in Table~\ref{table} are precise at a level of 1\% or better.
The magnitude and precision of the fitted line centers and widths are unaffected by amplitude scaling.

The ammonia spectral features at $v_{\rm LSR} \approx -50$, $-49$, and $-47\,{\rm km}\,{\rm s}^{-1}$ represent the first maser species in NGC~7538, besides water, to be observed having $v_{\rm LSR} > -51\,{\rm km}\,{\rm s}^{-1}$.
These weak features are not necessarily associated with IRS~1.
At 18~GHz, the GBT has a $-33.0$-dB sidelobe at the location of IRS~9 when pointed at IRS~1.
For example, the 20-mJy signal in Figure~\ref{gbt} may be a 40-Jy ammonia maser in IRS~9.
A large amplitude for an ammonia maser is not without precedent: Madden et al.\ (1986) and Pratap et al.\ (1991) find the (9,6) maser in W51 to have a flux density greater than 50~Jy.
IRS~11 lies in a null of the 18-GHz beam pattern and likely does not contribute to the observed signal at 18.5~GHz.
We are planning EVLA observations in order to precisely image the locations of all of the ammonia spectral features.

Urquhart et al.\ (2011) have described observations of NGC~7538 IRS~1-3 as part of the RMS survey of over 1000 massive young stellar objects.\footnote{http://www.ast.leeds.ac.uk/cgi-bin/RMS/RMS\_DATABASE.cgi}
The corresponding MSX source name is G111.5423+00.7776.
The survey includes GBT observations of the metastable $^{14}$NH$_3$ (1,1), (2,2), and (3,3) lines as well as the $6_{1,6} - 5_{2,3}$ water masers in the present study.
The RMS observations of G111.5423+00.7776 were conducted approximately 12~months prior to the current observations.
The telescope pointing position in the RMS survey is within an arcminute of the pointing position of the current data, meaning that the 22-GHz beam pattern in Figure~\ref{6cm}b is an accurate description of the RMS sky response.
G111.5423+00.7776 is one of only nine sources in their sample to exhibit $^{14}$NH$_3$ (1,1) emission at more than one velocity.
They find at least five broad ($\approx 3\,{\rm km}\,{\rm s}^{-1}$) thermal emission lines at $v_{\rm LSR} = -76.3$, $-64.0$, $-56.9$, $-48.9$, and $-37.2\,{\rm km}\,{\rm s}^{-1}$, consistent with the single-dish and interferometric spectra presented by Zheng et al.\ (2001).
Their (1,1) data are shown in the bottom panel of our Figure~\ref{gbt}.
We note three important distinctions between the metastable and nonmetastable ammonia data sets:
(1) there is no (1,1) emission at the velocity $v_{\rm LSR} = -60\,{\rm km}\,{\rm s}^{-1}$ of the strongest and longest-lived (9,6) masers, although there is (2,2) and (3,3) absorption, 
(2) in contrast, the new groups of (9,6) emission features centered at $v_{\rm LSR} = -56.9$ and $-48.9\,{\rm km}\,{\rm s}^{-1}$ each have corresponding (1,1) emission features, and
(3) the thermal (1,1), (2,2), and (3,3) emission lines are much broader than the (9,6) lines, suggestive of a nonthermal origin for the (9,6) emission, including the low-amplitude (9,6) features having $v_{\rm LSR} > -51\,{\rm km}\,{\rm s}^{-1}$.

\begin{figure}[h]
%\epsscale{.80}
\centering
\includegraphics*[angle=270,scale=0.35]{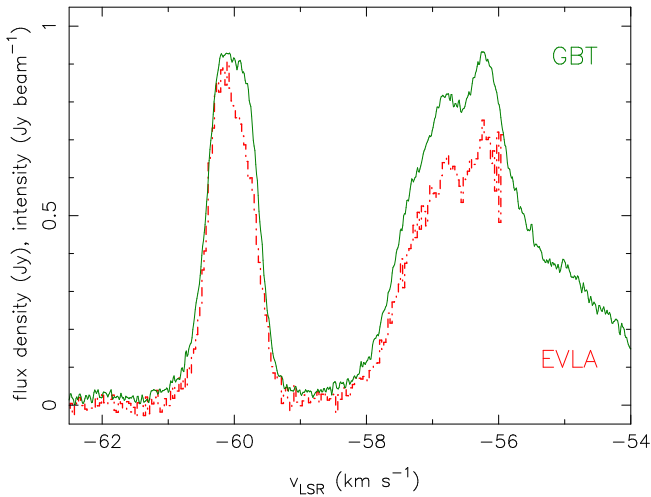}
%\plotone{spectrum.eps}
\caption{
Subset of the GBT spectrum of the $^{14}$NH$_3$ $(9,6)$ masers in NGC~7538 from 2011 January 22 (green line) overlaid with the $^{14}$NH$_3$ $(9,6)$ spectrum from the EVLA `DnC'-array observations from 2010 September 19 (red dash-dot histogram) from Hoffman \& Kim (Paper~I).
\label{evla}}
\end{figure}

The stronger (9,6) emission features near $v_{\rm LSR} = -60\,{\rm km}\,{\rm s}^{-1}$ and $-56\,{\rm km}\,{\rm s}^{-1}$ are known to arise in IRS~1 from the interferometric imaging presented in Paper~I.
The `SE' and `NW' labels from Paper~I are included in Table~\ref{table}.
In Figure~\ref{evla} is shown the GBT and EVLA spectra after subtraction of the continuum emission.
No significant differences are apparent between the observations aside from a multiplicative scaling factor comparable to calibration uncertainties.
The largest angular scale to which the EVLA observations were sensitive is approximately one minute of arc (approximately 1~pc at the distance of NGC~7538).
It is unlikely that there are masering regions of this size in NGC~7538 (see Fig.~\ref{6cm}) to which the interferometric observations could have been insensitive.

\subsection{H$_2$O and Comparisons to NH$_3$\label{waterResults}}

\begin{figure*}
%\epsscale{.80}
\centering
\includegraphics*[angle=270,scale=0.67]{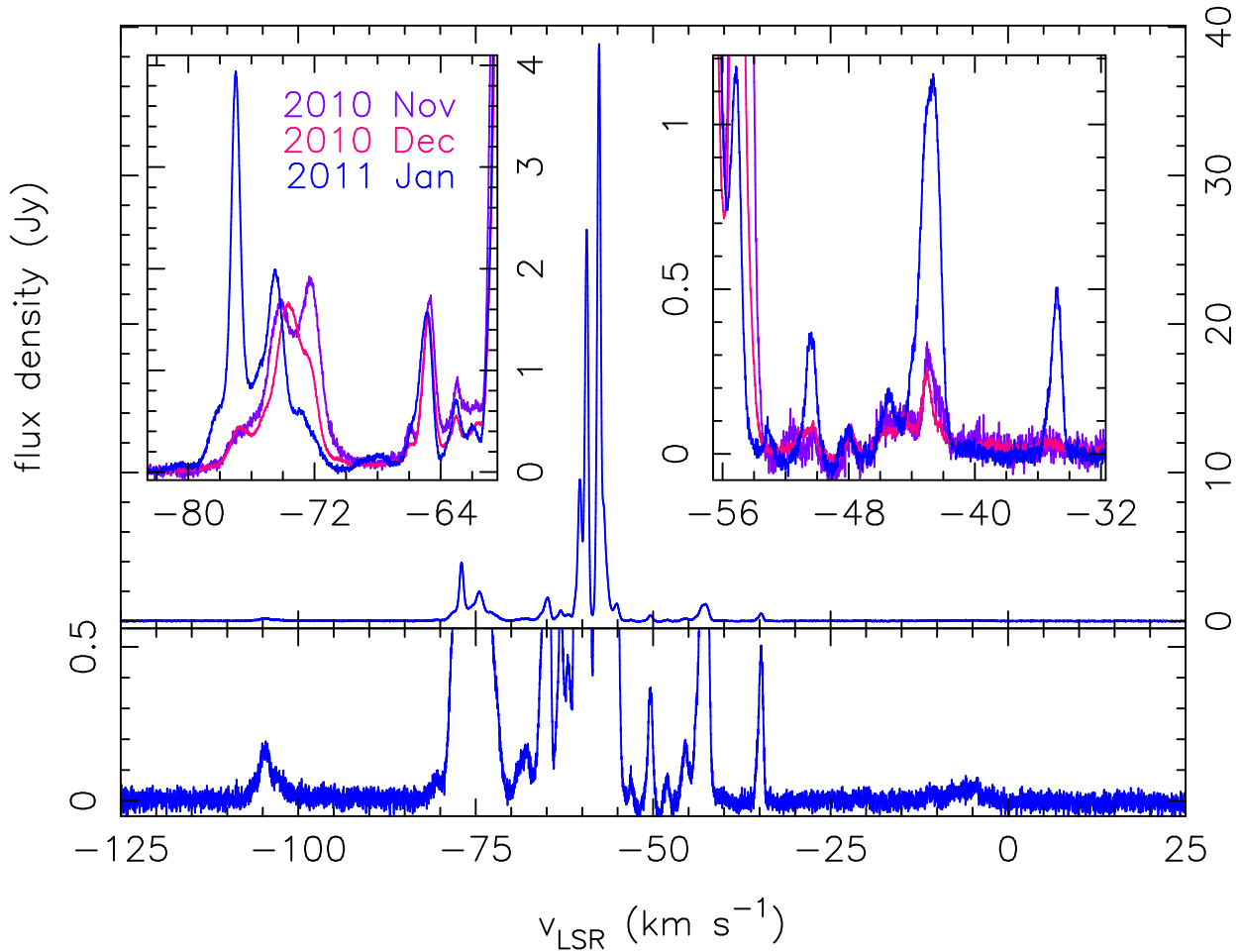}
%\plotone{spectrum.eps}
\caption{
GBT spectrum of the H$_2$O masers in NGC~7538 from 2011 January 22 (blue line throughout) and 2010 December 8 and 2010 November 24 (pink and purple lines in insets).
The velocities and amplitudes of all of the emission features from all three epochs are summarized in Table~\ref{waterTable}.
The bottom panel is a magnified view of the lowest amplitudes, covering the same velocity range as the top panel.
The insets have the same horizontal velocity scale as each other and include data from all three epochs at intermediate velocities; the data displayed in the insets is representative of typical variability and of the uncertainty in the assignment of velocity features among the epochs.
\label{h2o}}
\end{figure*}

The emission from water masers in each of the three epochs is shown in Figure~\ref{h2o}.
Among the epochs, we observed intensity variability typical of water masers.
The amplitudes of the features at each epoch are summarized in Table~\ref{waterTable}.
We did not detect Zeeman splitting between the right- and left-circular polarizations in any of the water masers (see also, Surcis et al.\ 2011).

\begin{deluxetable}{l l l l}
\tablecolumns{4}
\tablewidth{0pt}
\tablecaption{Fitted Spectral Properties of Water Emission\label{waterTable}}
\tablehead{
\multicolumn{1}{c}{$v_{\rm LSR}$} & \multicolumn{1}{c}{$S_{\rm Nov}$\tablenotemark{$\dagger$}} & \multicolumn{1}{c}{$S_{\rm Dec}$\tablenotemark{$\dagger$}} & \multicolumn{1}{c}{$S_{\rm Jan}$\tablenotemark{$\dagger$}} \\
\multicolumn{1}{c}{(${\rm km}\,{\rm s}^{-1}$)} & \multicolumn{1}{c}{(Jy)} & \multicolumn{1}{c}{(Jy)} & \multicolumn{1}{c}{(Jy)}
}
\startdata
$-$105.1(1)  & 0.14(1)  & 0.06(1) & 0.15(1)  \\
$-$104.4(1)  & 0.12(6)  & 0.06(1) & 0.17(1)  \\
$-$103.0(2)  & $<$0.05  & $<$0.03 & 0.08(1)  \\
$-$96.7(3)   & 0.05(3)  & 0.04(2) & 0.03(2)  \\
$-$80.4(2)   & $<$0.05  & $<$0.03 & 0.07(2)  \\
$-$78.0(2)   & $<$0.05  & $<$0.03 & 0.65(2)  \\
$-$77.0(2)   & 0.34(2)  & 0.40(2) & 3.92(4)  \\
$-$75.4(3)\tablenotemark{a}   & $<$0.05  & 0.5(1)   & 1.0(1)   \\ % X
$-$74.5(1)\tablenotemark{a}   & $<$0.05  & 0.10(5)  & 1.95(5)   \\ % 9
$-$74.1(2)\tablenotemark{a}   & 0.5(1)   & $<$0.03  & $<$0.03   \\
$-$73.5(2)\tablenotemark{a}   & $<$0.05  & 1.6(1)   & $<$0.03   \\
$-$72.8(2)\tablenotemark{a}   & 0.10(8)  & $<$0.03  & 0.12(5)   \\
$-$72.2(2)   & 1.8(2)   & 0.15(3)  & 0.10(5)   \\
$-$68.0(7)   & $<$0.05  & $<$0.03  & 0.17(3)   \\
$-$67.0(3)   & 0.15(5)  & $<$0.03  & $<$0.05   \\ % 11
$-$65.9(1)   & 0.44(4)  & 0.26(3)  & 0.05(3)   \\
$-$64.8(1)\tablenotemark{a}   & 1.70(4)  & 1.52(2)  & 1.56(2)   \\
$-$63.0(1)   & 0.88(3)  & 0.52(3)  & 0.67(3)   \\ % F
$-$62.0(1)   & 0.3(1)   & 0.3(1)   & 0.3(1)    \\ % 123, 11
$-$60.36(4)  & 8.8(1)   & 5.4(1)   & 8.9(1)    \\ % X
$-$59.38(1)  & 22.8(4)  & 15.3(2)  & 25.9(2)   \\ % 11
$-$58.53(4)  & 4.6(2)   & 1.6(3)   & $<$0.03   \\
$-$57.65(1)  & 49.2(3)  & 29.8(1)  & 38.3(1)   \\ % G
$-$56.9(2)   & 3.4(5)   & 0.8(5)   & 2.2(5)    \\ % X, 11
$-$55.02(1)  & 9.1(2)   & 4.2(1)   & 0.7(2)    \\
$-$54.47(2)  & 4.7(3)   & $<$0.1   & $<$0.1    \\
$-$53.10(5)  & $<$0.05  & $<$0.03  & 0.08(2)   \\
$-$51.7(1)   & 0.06(2)  & 0.03(2)  & $<$0.03   \\
$-$50.40(1)  & 0.055(8) & 0.08(2)  & 0.37(1)   \\
$-$47.95(5)  & 0.06(1)  & 0.06(1)  & 0.06(1)   \\
$-$45.46(5)  & 0.06(4)  & 0.05(3)  & 0.17(2)   \\ % 123
$-$44.0(1)   & $<$0.05  & $<$0.03  & 0.05(2)   \\
$-$43.01(2)  & 0.31(6)  & 0.24(4)  & 0.92(2)   \\
$-$42.45(1)  & 0.1(5)   & $<$0.1   & 0.46(4)   \\
$-$38.88(6)  & 0.03(2)  & 0.04(1)  & $<$0.03   \\
$-$34.84(1)  & 0.05(3)  & 0.03(1)  & 0.48(1)   \\
$-$26.0(1)   & 0.05(2)  & 0.03(2)  & $<$0.03   \\
$-$21.6(1)   & $<$0.05  & $<$0.03  & 0.02(1)   \\
$-$19.9(1)   & $<$0.05  & $<$0.03  & 0.02(1)   \\
$-$12.0(1)   & 0.04(2)  & 0.03(1)  & $<$0.03   \\
$-$10.94(7)  & $<$0.05  & 0.03(2)  & 0.03(1)   \\
$-$ 7.8(1)   & $<$0.05  & $<$0.03  & 0.04(1)   \\
$-$ 6.4(1)   & 0.05(3)  & 0.03(2)  & 0.04(1)   \\
$-$ 4.2(1)   & $<$0.05  & $<$0.03  & 0.05(1)   \\ [-3mm]
\enddata
\tablecomments{The number in parentheses is the uncertainty in the final digit.}
\tablenotetext{$\dagger$}{The absolute intensity calibration of the observations is precise within $\approx 10$\% while the higher precision quoted here is appropriate for relative comparisons among the features in an epoch.}
\tablenotetext{a}{Inter-epoch relationship of features in this row is not certain.}
\end{deluxetable}

In a series of papers based on single-dish observations of the water masers in NGC~7538, Lekht et al.\ (2003; 2004; 2007; 2010) have found 0.9-year and 13-year periodicities in the intensity variability.
For our observation dates in late 2010, the phase of their models is at its lowest amplitude.
Indeed, the integrated fluxes we measure in our three epochs -- $85$, $51$, and $68\,{\rm Jy}\,{\rm km}\,{\rm s}^{-1}$ -- are consistent with their lowest measurement of $50\,{\rm Jy}\,{\rm km}\,{\rm s}^{-1}$ (compared with values of $\approx 800\,{\rm Jy}\,{\rm km}\,{\rm s}^{-1}$ at maximum phase).
We find no monotonic or correlated variability among the four maser velocities studied by Lekht et al.: $-61.5$, $-60$, $-58$, and $-54.5\,{\rm km}\,{\rm s}^{-1}$.
Additionally, timescales of variability as short as days have been observed for water masers, in general (e.g., Liljestrom et al.\ 1989; Zhou \& Zheng 2001).
Our three epochs of observations do not have sufficient temporal coverage for comparisons with models of short-term variability.

As with the ammonia emission discussed previously, not all of the water emission apparent in Figure~\ref{h2o} is necessarily associated with IRS~1.
Emission from water masers is observed from IRS~9 and IRS~11 as well as regions unassociated with infrared emission, denoted by letters and circles in Figure~\ref{6cm}b.
For example, the 150-mJy feature we observe at $v_{\rm LSR} = -67.0\,{\rm km}\,{\rm s}^{-1}$ is likely the strong ($\approx 50$~Jy) maser observed in IRS~11 by Kameya et al.\ (1990) in the second sidelobe of our GBT pointing.

We detect water emission at velocities well removed from the systemic velocity of NGC~7538 of $v_{\rm LSR} \approx -58\,{\rm km}\,{\rm s}^{-1}$.
At the extremes, weak emission is observed at $-105\,{\rm km}\,{\rm s}^{-1}$ and at $-4\,{\rm km}\,{\rm s}^{-1}$. 
Emission has been observed from IRS~11 at velocities as low as $-146\,{\rm km}\,{\rm s}^{-1}$ by Sunada et al.\ (2007), but they did not observe strong emission near $-105\,{\rm km}\,{\rm s}^{-1}$ nor near $-4\,{\rm km}\,{\rm s}^{-1}$.
With a comparable pointing of the GBT, Urquhart et al.\ (2011) observed the range $-137\,{\rm km}\,{\rm s}^{-1} < v_{\rm LSR} < 22\,{\rm km}\,{\rm s}^{-1}$ with a sensitivity of approximately 150~mJy and detect the feature at $-105\,{\rm km}\,{\rm s}^{-1}$ but not the feature at $-4\,{\rm km}\,{\rm s}^{-1}$. 
No other authors have reported water emission from anywhere in NGC~7538 with velocity $v_{\rm LSR} > -38\,{\rm km}\,{\rm s}^{-1}$ (Urquhart et al.\ 2011).
If these masers lie in IRS~11 in a sidelobe of our observations, their actual flux densities would be $\sim 100$~Jy, which is an atypically large amplitude for a water maser at these outlying velocities.

The only water maser known to occur at a velocity and position comparable to a $^{14}$NH$_3$ (9,6) maser is at $v_{\rm LSR} = -56.9\,{\rm km}\,{\rm s}^{-1}$ (Paper~I; Kameya et al.\ 1990; Galv\'an-Madrid et al.\ 2010).
This water maser exhibits a four-fold intensity fluctuation during our observations while the corresponding ammonia maser exhibits no variability.
If these two maser species are associated with the same physical volume (for example, a molecular clump) then their pumping mechanisms or saturation states must be substantially different from each other.

In comparing our results with other maser studies, we classify the masers in NGC~7538 into three groups:
(1) those in the narrow velocity range $-61\,{\rm km}\,{\rm s}^{-1} < v_{\rm LSR} < -51\,{\rm km}\,{\rm s}^{-1}$ having positions near the continuum peak of IRS~1, presumably related to the inner regions of a circumstellar disk and torus, including rare species such as 23.1-GHz CH$_3$OH and 4.8-GHz H$_2$CO (Galv\'an-Madrid et al.\ 2010; Hoffman et al.\ 2003),
(2) the OH masers in the same narrow velocity range but with positions relatively far from the continuum peak and unassociated with the central disk structure (Hutawarakorn \& Cohen 2003), and
(3) the H$_2$O and $^{14}$NH$_3$ masers that have velocities outside of the narrow range.
Since we find no significant correlated velocities or variability between the water and ammonia masers, it remains an open question what relationship, if any, exists between water and ammonia.
The drawing of further conclusions about the physical conditions illuminated by the ammonia sources awaits interferometric imaging of the new ammonia masers.

In the case of the water masers, interferometric positions are already known from Kameya et al.\ (1990).
They find the water maser at $-46.4\,{\rm km}\,{\rm s}^{-1}$ to lie within IRS~1, but the remainder of the high-velocity masers at $-48.5\,{\rm km}\,{\rm s}^{-1}$ and $-45.4\,{\rm km}\,{\rm s}^{-1}$ have positions well removed from any other infrared source or maser (see their Figure~3), although still in the main beam of our GBT observations.
Qiu et al.\ (2011) have recently reported cores of millimeter-wave emission (MM2 and MM3) at the positions of the $-48.5\,{\rm km}\,{\rm s}^{-1}$ and $-45.4\,{\rm km}\,{\rm s}^{-1}$ water masers.
The MM2 and MM3 sources are interpreted as embedded stars that are comparable in gas mass to the power source of IRS~1, presumably capable of bearing their own circumstellar maser systems.

Most models of the IRS~1 region (e.g., Lekht et al.\ 2007; Surcis et al.\ 2011) include only the masers in the velocity range $-61\,{\rm km}\,{\rm s}^{-1} < v_{\rm LSR} < -51\,{\rm km}\,{\rm s}^{-1}$.
The highest velocity masers in these models have been accelerated by infall toward the central star and are therefore located close to the peak of continuum emission in IRS~1.
The high-velocity (i.e., $v_{\rm LSR} > -51\,{\rm km}\,{\rm s}^{-1}$) occurrences of water and ammonia masers may instead be associated with either different cores or with outflow interaction sites having no associated continuum emission.
Indeed, higher velocity water masers appear to be a different class than the luminous water masers that are associated with OH masers near the systemic velocity (Liu et al.\ 2007).
We are planning interferometric imaging of the new ammonia masers in order to determine precise locations with respect to the IRS and MM sources.

\subsection{CCS}

We searched for thermal emission from the CC$^{32}$S and CC$^{34}$S molecules using two spectral windows centered on the $J_N = 2_1 - 1_0$ transition.
In summing the entire on-source time for all three epochs (2.15~hours), we do not detect either molecule at an {\it rms} noise level of 8~mJy (18~mK) in a single channel.

The CCS molecule is the subject of ongoing study as an observational diagnostic of the age and chemistry of molecular cloud cores (see, for example, Roy et al.\ 2011).
The current observations represent the most sensitive search to date for either the CC$^{32}$S or CC$^{34}$S molecule.

The CCS molecule is suggested to exist only in the early stages of molecular cloud evolution, having a lifetime $\simeq (0.7-3) \times 10^4$~yr (de Gregorio-Monsalvo et al.\ 2006).
Most pertinently, CCS is not expected to coexist with NH$_3$: it is indicated by models that CCS is destroyed in cores that have evolved to an age at which NH$_3$ can be generated (Suzuki et al.\ 1992).
The age of IRS~1 has been modelled from various spectral line observations by Knez et al.\ (2009) to be in the range $10^{3-6}$~yr.
Qualitatively, Qiu et al.\ (2011) suggest that source MM2 is the youngest core in the NGC~7538 complex.
An independent, qualitative estimate of age comes from Hutawarakorn \& Cohen (2003) who model the polarization of OH masers in NGC~7538 and suggest that IRS~1 is older than IRS~9 and IRS~11, and therefore more likely to bear ammonia.
Our CCS results are suggestive of an age greater than $\approx 10^4$~yr, not only for IRS~1 but for every source in the main beam of our observations, including MM2 and MM3.

\section{Conclusions}

Using the single-dish Green Bank Telescope, we report new maser emission at 18.5~GHz from $^{14}$NH$_3$ (9,6) in NGC~7538.
Maser emission from ammonia has been observed over the velocity range $-60\,{\rm km}\,{\rm s}^{-1} < v_{\rm LSR} < -46\,{\rm km}\,{\rm s}^{-1}$, making ammonia and water the only maser species yet observed having $v_{\rm LSR} > -51\,{\rm km}\,{\rm s}^{-1}$.
No variability is detected in the ammonia masers in three observing epochs that span a timescale of two months.
Water maser emission is observed over the velocity range $-105\,{\rm km}\,{\rm s}^{-1} < v_{\rm LSR} < -4\,{\rm km}\,{\rm s}^{-1}$, extending the highest known velocity in NGC~7538 from the previous high of $-38\,{\rm km}\,{\rm s}^{-1}$.
We do not detect either CC$^{32}$S or CC$^{34}$S in the most sensitive search to date toward any source for these molecules, indicating an age greater than $\approx 10^4$~yr for the IRS~1-3 region.

\acknowledgments

We thank J.\ S.\ Urquhart for communicating the RMS survey results prior to publication.
We thank D.\ A.\ Roshi and the Green Bank staff for support during this project.
We are grateful for a careful reading of our manuscript by an anonymous referee.
The authors acknowledge support from the Thomas Penrose Bennett Prize Fund and the Lovejoy Science Fund of St.\ Paul's School.

{\it Facilities:} \facility{GBT ()}


\begin{thebibliography}{}
\bibitem[Akabane et al.\ (1992)]{aka92} Akabane, K., Tsunekawa, S., Inoue, M., Kawabe, R., Ohashi, N., Kameya, O., Ishiguro, M., Sofue, Y.\ 1992, \pasj, 44, 421
%\bibitem[Araya et al.\ (2007)]{ara07} Araya, E., Hofner, P., Goss, W.\ M., Linz, H., Kurtz, S., Olmi, L.\ 2007, \apjs, 170, 152
%\bibitem[Araya et al.\ (2009)]{ara09} Araya, E., Kurtz, S., Hofner, P., Linz, H.\ 2009, \apj, 698, 1321
%\bibitem[Breen et al.\ (2010)]{bre10} Breen, S.\ L., Caswell, J.\ L., Ellingsen, S.\ P., Phillips, C.\ J.\ 2010, \mnras, 406, 1487
%\bibitem[Brown \& Cragg (1991)]{bro91} Brown, R.\ D., Cragg, D.\ M.\ 1991, \apj, 378, 445
\bibitem[de Gregorio-Monsalvo et al.\ (2006)]{deg06} de Gregorio-Monsalvo, I., G\'omez, J.\ F., Su\'arez, O., Kuiper, T.\ B.\ H., Rodr\'iguez, L.\ F., Jim\'enez-Bail\'on, E.\ 2006, \apj, 642, 319
\bibitem[Galv\'an-Madrid et al.\ (2010)]{gal10} Galv\'an-Madrid, R., Montes, G., Ramírez, E.\ A., Kurtz, S., Araya, E., Hofner, P.\ 2010, \apj, 713, 423
%\bibitem[Gaume et al.\ (1991)]{gau91} Gaume, R.\ A., Johnston, K.\ J., Nguyen, H.\ A., Wilson, T.\ L., Dickel, H.\ R., Goss, W.\ M., Wright, M.\ C.\ H.\ 1991, \apj, 376, 608
%\bibitem[Gaume et al.\ (1993)]{gau93} Gaume, R.\ A., Johnston, K.\ J., Wilson, T.\ L.\ 1993, \apj, 417, 645
%\bibitem[Gibb et al.\ (2001)]{gib01} Gibb, E.\ L., Whittet, D.\ C.\ B., Chiar, J.\ E.\ 2001, \apj, 558, 702
\bibitem[Hoffman et al.\ (2003)]{H03} Hoffman, I.\ M., Goss, W.\ M., Palmer, P., Richards, A.\ M.\ S.\ 2003, \apj, 598, 1061
\bibitem[Hoffman \& Kim (2011)]{hof11} Hoffman, I.\ M.\ \& Kim, S.\ S.\ 2011, \apjl, 739, L15 (Paper~I)
%\bibitem[Hoffman et al.\ (2007)]{H07} Hoffman, I.\ M., Goss, W.\ M., Palmer, P.\ 2007, \apj, 654, 971
\bibitem[Hutawarakorn \& Cohen (2003)]{hut03} Hutawarakorn, B., Cohen, R.\ J.\ 2003, \mnras, 345, 175
%\bibitem[Johnston et al.\ (1989)]{joh89} Johnston, K.\ J., Stolovy, S.\ R., Wilson, T.\ L., Henkel, C., Mauersberger, R.\ 1989, \apjl, 343, L41
\bibitem[Kameya et al.\ (1990)]{kam90} Kameya, O., Morita, K., Kawabe, R., Ishiguro, M.\ 1990, \apj, 355, 562
\bibitem[Kawabe et al.\ (1992)]{kaw92} Kawabe, R., Suzuki, M., Hirano, N., Akabane, K., Barsony, M., Najita, J.\ R., Kameya, O., Ishiguro, M.\ 1992, \pasj, 44, 435
\bibitem[Knez et al.\ (2009)]{kne09} Knez, C., Lacy, J.\ H., Evans, N.\ J., van Dishoeck, E.\ F., Richter, M.\ J.\ 2009, \apj, 696, 471
\bibitem[Lekht et al.\ (2003)]{lek03} Lekht, E.\ E., Munitsyn, V.\ A., Tolmachev, A.\ M.\ 2003 Astron.\ Rep.\ 47, 838
\bibitem[Lekht et al.\ (2004)]{lek04} Lekht, E.\ E., Munitsyn, V.\ A., Tolmachev, A.\ M.\ 2004 Astron.\ Rep.\ 48, 200
\bibitem[Lekht et al.\ (2007)]{lek07} Lekht, E.\ E., Munitsyn, V.\ A., Krasnov, V.\ V.\ 2007 Astron.\ Rep.\ 51, 27
\bibitem[Lekht et al.\ (2010)]{lek10} Lekht, E.\ E., Munitsyn, V.\ A.\ 2010 Astron.\ Rep.\ 54, 151
\bibitem[Liljestrom et al.\ (1989)]{lil89} Liljestrom, T., Mattila, K., Toriseva, M., Anttila, R.\ 1989, \aaps, 79, 19
\bibitem[Liu et al.\ (2007)]{liu07} Liu, H., Forster, J.\ R., Liu, Y., Sun, J.\ 2007, \apss, 307, 357
\bibitem[Madden et al.\ (1986)]{mad86} Madden, S.\ C., Irvine, W.\ M., Matthews, H.\ E., Brown, R.\ D., \& Godfrey, P.\ D.\ 1986, \apjl, 300, L79
%\bibitem[Mauersberger et al.\ (1987)]{mau87} Mauersberger, R., Henkel, C., Wilson, T.\ L.\ 1987, \aap, 173, 352
\bibitem[Moscadelli et al.\ (2009)]{mos09} Moscadelli, L., Reid, M.\ J., Menten, K.\ M., Brunthaler, A., Zheng, X.\ W., Xu, Y.\ 2009, \apj, 693, 406
%\bibitem[Pratap et al.\ (1989)]{pra89} Pratap, P., Batrla, W., Snyder, L.\ E.\ 1989, \apj, 341, 832
\bibitem[Pratap et al.\ (1991)]{pra91} Pratap, P., Menten, K.\ M., Reid, M.\ J., Moran, J.\ M., Walmsley, C.\ M.\ 1991, \apjl, 373, L13
\bibitem[Price et al.\ (2001)]{pri01} Price, S.\ D., Egan, M.\ P., Carey, S.\ J., Mizuno, D.\ R., Kuchar, T.\ A.\ 2001, \aj, 121, 2819
\bibitem[Qiu et al.\ (2011)]{qui11} Qiu, K., Zhang, Q., Menten, Karl M.\ 2011, \apj, 728, 6
%\bibitem[Rots et al.\ (1981)]{rot81} Rots, A.\ H., Dickel, H.\ R., Forster, J.\ R., Goss, W.\ M.\ 1981, \apjl, 245, L15
\bibitem[Roy et al.\ (2011)]{roy11} Roy, N., Datta, A., Momjian, E., Sarma, A.\ P.\ 2011, \apjl, 739, L4
\bibitem[Sandell et al.\ (2009)]{san09} Sandell, G., Goss, W.\ M., Wright, M., \& Corder, S.\ 2009, \apjl, 699, L31
\bibitem[Srikanth (2009)]{sri09} Srikanth, S.\ 2009, GBT Memo \#262
\bibitem[Sunada et al.\ (2007)]{sun07} 	Sunada, K., Nakazato, T., Ikeda, N., Hongo, S., Kitamura, Y., Yang, J.\ 2007, \pasj, 59, 1185
\bibitem[Surcis et al.\ (2011)]{sur11} 	Surcis, G., Vlemmings, W.\ H.\ T., Torres, R.\ M., van Langevelde, H.\ J., Hutawarakorn Kramer, B.\ 2011, \aap, 533, A47
\bibitem[Suzuki et al.\ (1992)]{suz92} Suzuki, H., Yamamoto, S., Ohishi, M., Kaifu, N., Ishikawa, S., Hirahara, Y., Takano, S.\ 1992, \apj, 392, 551
\bibitem[Urquhart et al.\ (2011)]{urq11} Urquhart, J.\ S., Morgan, L.\ K., Figura, C.\ C., Moore, T.\ J.\ T., Lumsden, S.\ L., Hoare, M.\ G., Oudmaijer, R.\ D., Mottram, J.\ C., Davies, B., Dunham, M.\ K.\ 2011, \mnras, in press, arXiv:1107.3913
\bibitem[Werner et al.\ (1979)]{wer79} Werner, M.\ W., Becklin, E.\ E., Gatley, I., Matthews, K., Neugebauer, G., Wynn-Williams, C.\ G.\ 1979, \mnras, 188, 463
%\bibitem[Wilson et al.\ (1984)]{wil84} Wilson, T.\ L., Walmsley, C.\ M., Snyder, L.\ E., Jewell, P.\ R.\ 1984, \aap, 134, L7
\bibitem[Wilson \& Henkel (1988)]{wil88} Wilson, T.\ L., Henkel, C.\ 1988, \aap, 206, L26
\bibitem[Wilson et al.\ (1990)]{wil90} Wilson, T.\ L., Henkel, C., Johnston, K.\ J.\ 1990, \aap, 229, L1
%\bibitem[Wynn-Williams et al.\ (1974)]{wyn74} Wynn-Williams, C.\ G., Becklin, E.\ E., Neugebauer, G.\ 1974, \apj, 187, 473
\bibitem[Zheng et al.\ (2001)]{zhe01} Zheng, X.-W., Zhang, Q., Ho, P.\ T.\ P., Pratap, P.\ 2001, \apj, 550, 301
\bibitem[Zhou \& Zheng (2001)]{zho01} Zhou, J., Zheng, X.\ 2001, \apss, 275, 431
\end{thebibliography}
\end{document}